\documentclass[twocolumn,showpacs,preprintnumbers,amsmath,amssymb]{revtex4}
\usepackage{graphicx}
\usepackage{dcolumn}
\usepackage{bm}

\begin{document}

\title{Low-lying dipole resonance in neutron-rich Ne isotopes
}

\author{Kenichi Yoshida$^{1,2}$}
\author{Nguyen Van Giai$^{2}$}
\affiliation{
$^{1}$Department of Physics, Graduate School of Science, Kyoto University, 
Kyoto 606-8502, Japan
\\
$^{2}$Institut de Physique Nucl\'eaire, IN$_{2}$P$_{3}$-CNRS, 
and Universit\'e Paris-Sud, F-91406 Orsay Cedex, France
}%

\date{\today}

\begin{abstract}
Microscopic structure of the low-lying isovector dipole excitation mode 
in neutron-rich $^{26,28,30}$Ne is investigated by performing 
deformed quasiparticle-random-phase-approximation (QRPA) calculations. 
The particle-hole residual interaction is derived from a Skyrme force 
through a Landau-Migdal approximation.  
We have obtained the low-lying resonance in $^{26}$Ne at around 8.5 MeV. 
It is found that the isovector dipole strength at $E_{x}<10$ MeV exhausts 
about 6.0\% of the classical Thomas-Reiche-Kuhn dipole sum rule. 
This excitation mode is composed of several QRPA eigenmodes, 
one is generated by a $\nu(2s^{-1}_{1/2} 2p_{3/2})$ transition dominantly, 
and the other mostly by a $\nu(2s^{-1}_{1/2} 2p_{1/2})$ transition.
The neutron excitations take place outside of the nuclear surface 
reflecting the spatially extended structure of the $2s_{1/2}$ wave function. 
In $^{30}$Ne, the deformation splitting of the giant resonance is large, and 
the low-lying resonance is overlapping with the giant resonance.  
\end{abstract}

\maketitle

\section{Introduction}
The study of nuclei far off stability is one of the 
most active research fields in nuclear physics~\cite{tan01,hor01,hag02a}, 
and exploring the collective motions unique in unstable nuclei is one of 
the main issues experimentally and theoretically~\cite{neu07}. 
In neutron-rich nuclei, because of the absence of the Coulomb barrier 
the surface structure is quite different from stable nuclei. 
One of the unique structures is the neutron skin~\cite{suz95,miz00}. 
Since the collective excitations are sensitive to the surface structure, 
one can expect new kinds of exotic excitation modes associated with the neutron skin 
to appear in neutron-rich nuclei. 
One of the examples is the soft dipole excitation~\cite{ike88}, 
which is observed not only 
in light halo nuclei~\cite{sac93,shi95,zin97,nak06,nak94,pal03,fuk04,nak99,pra03,aum99}, 
but also in heavier systems~\cite{lei01,try03,adr05}, where an appreciable 
$E1$ strength is observed above the neutron threshold 
exhausting several percents of the energy-weighted sum rule (EWSR). 

The structure of the low-lying dipole state and its collectivity 
has been studied in the framework of the mean-field calculations 
by many groups~\cite{cat97,ham98,ham99,gor02,mat02,ter06,col01,sar04,vre01a,vre01b,paa05,cao05,liv07}.
A low-lying dipole state in neutron-rich $^{26}$Ne was first predicted by using the 
relativistic quasiparticle-random-phase approximation (QRPA) in Ref.~\cite{cao05}, 
and recently it was observed at RIKEN around 9 MeV, exhausting 
about 5\% of the Thomas-Reiche-Kuhn (TRK) dipole sum rule~\cite{gib07}. 
In Ref.~\cite{cao05}, the QRPA was solved in the response function formalism. 
This method can treat the excitations to the continuum exactly by employing the 
Green's functions satisfying the out-going-wave boundary conditions, 
but an additional procedure is required to obtain the microscopic structure 
of the excitation mode~\cite{kha05}. 

In the present paper, we investigate the microscopic structure of the low-lying 
dipole resonance in neutron-rich Ne isotopes, 
and we discuss the isotopic dependence with special attention to the deformation effects. 
To this end, 
we have developed a deformed QRPA code in the matrix formulation 
based on the coordinate-space Skyrme-Hartree-Fock-Bogoliubov (HFB) theory. 

The paper is organized as follows: 
In Sec.~\ref{model}, we explain our method. 
In Sec.~\ref{check}, we check the results of our new calculation scheme by comparing the 
existing QRPA results. 
In Sec.~\ref{results}, we present the results of the deformed QRPA and we discuss the 
microscopic structure of the low-lying dipole state in $^{26,28,30}$Ne. 
Finally, we summarize the paper in Sec.~\ref{summary}.

\section{\label{model}Model}
We briefly summarize here our approach (see Ref.~\cite{yos06} for details).  
In order to discuss simultaneously effects of nuclear deformation 
and pairing correlations including the continuum, 
we solve the HFB equations~\cite{dob84,bul80}
\begin{multline}
\begin{pmatrix}
h^{\tau}(\boldsymbol{r}\sigma)-\lambda^{\tau} & \tilde{h}^{\tau}(\boldsymbol{r}\sigma) \\
\tilde{h}^{\tau}(\boldsymbol{r}\sigma) & -(h^{\tau}(\boldsymbol{r}\sigma)-\lambda^{\tau}) \end{pmatrix}
\begin{pmatrix}
\varphi^{\tau}_{1,\alpha}(\boldsymbol{r}\sigma) \\ 
\varphi^{\tau}_{2,\alpha}(\boldsymbol{r}\sigma)
\end{pmatrix}
\\
= E_{\alpha}
\begin{pmatrix}
\varphi^{\tau}_{1,\alpha}(\boldsymbol{r}\sigma) \\ 
\varphi^{\tau}_{2,\alpha}(\boldsymbol{r}\sigma)
\end{pmatrix} \label{eq:HFB1}
\end{multline}
directly in the cylindrical coordinates 
assuming axial and reflection symmetries. 
Here, $\tau=\nu$ (neutron) and $\pi$ (proton), and $\boldsymbol{r}=(\rho,z,\phi)$. 
For the mean-field Hamiltonian $h$, 
we employ the SkM* interaction~\cite{bar82}.
Details for expressing the densities and currents in the cylindrical coordinate representation 
can be found in Refs.~\cite{ter03,sto05}. 
The pairing field is treated by using 
the density-dependent contact interaction~\cite{ber91,ter95}, 
\begin{equation}
v_{pp}(\boldsymbol{r},\boldsymbol{r}^{\prime})=V_{0}\dfrac{1-P_{\sigma}}{2}
\left[ 1-  \left(\dfrac{\varrho^{\mathrm{IS}}(\boldsymbol{r})}{\varrho_{0}}\right)^{\gamma} \right]
\delta(\boldsymbol{r}-\boldsymbol{r}^{\prime}). \label{eq:res_pp}
\end{equation}
with $V_{0}=-390$ MeV $\cdot$fm$^{2}$ and $\varrho_{0}=0.16$ fm$^{-3}$, $\gamma=1$. 
Here, $\varrho^{\mathrm{IS}}(\boldsymbol{r})$ denotes the isoscalar density and 
$P_{\sigma}$ the spin exchange operator. 
The pairing strength $V_{0}$ is determined so as to approximately reproduce the 
experimental pairing gap of 1.25 MeV in $^{28}$Ne obtained by the three-point formula~\cite{sat98}. 
Because the time-reversal symmetry and reflection symmetry 
with respect to the $x-y$ plane are assumed, 
we have only to solve for positive $\Omega$ and positive $z$. 
We use the lattice mesh size $\Delta\rho=\Delta z=0.6$ fm 
and the box boundary condition at  
$\rho_{\mathrm{max}}=9.9$ fm and $z_{\mathrm{max}}=9.6$ fm. 
The quasiparticle energy is cut off at 60 MeV and 
the quasiparticle states up to $\Omega^{\pi}=13/2^{\pm}$ are included. 

Using the quasiparticle basis obtained by solving the HFB equation (\ref{eq:HFB1}),  
we solve the QRPA equation in the matrix formulation~\cite{row70} 
\begin{equation}
\sum_{\gamma \delta}
\begin{pmatrix}
A_{\alpha \beta \gamma \delta} & B_{\alpha \beta \gamma \delta} \\
B_{\alpha \beta \gamma \delta} & A_{\alpha \beta \gamma \delta}
\end{pmatrix}
\begin{pmatrix}
X_{\gamma \delta}^{\lambda} \\ Y_{\gamma \delta}^{\lambda}
\end{pmatrix}
=\hbar \omega_{\lambda}
\begin{pmatrix}
1 & 0 \\ 0 & -1
\end{pmatrix}
\begin{pmatrix}
X_{\alpha \beta}^{\lambda} \\ Y_{\alpha \beta}^{\lambda}
\end{pmatrix} \label{eq:AB1}.
\end{equation}
The residual interaction in the particle-particle (p-p) channel 
appearing in the QRPA matrices $A$ and $B$ is 
the density-dependent contact interaction (\ref{eq:res_pp}).
On the other hand, 
for the residual interaction in the particle-hole (p-h) channel, 
we employ the Landau-Migdal (LM) approximation~\cite{bac75} 
applied to the density-dependent Skyrme forces~\cite{gia81,gia98}, 
\begin{align}
v_{ph}(\boldsymbol{r},\boldsymbol{r}^{\prime})=&
N_{0}^{-1}\{F_{0}+F_{0}^{\prime}\tau\cdot\tau^{\prime} \notag \\
&+(G_{0}+G_{0}^{\prime} \tau\cdot\tau^{\prime})\sigma\cdot\sigma^{\prime} \}
\delta(\boldsymbol{r}-\boldsymbol{r}^{\prime}). \label{eq:res_ph}
\end{align}
Here, $N_{0}$ is the density of states and the Landau parameters are deduced from the same 
Skyrme force which generates the mean field. 
Because the full self-consistency between the static mean-field calculation 
and the dynamical QRPA calculation is broken, 
we have to renormalize the residual interaction in the particle-hole channel 
by an overall factor $f_{ph}$ to get the spurious $K^{\pi}=0^{-}$ or $1^{-}$  modes
(representing the center-of-mass motion) at zero energy
($v_{ph} \rightarrow f_{ph}\cdot v_{ph}$).  
We cut the two-quasiparticle space at $E_{\alpha}+E_{\beta} \leq 60$ MeV 
due to the excessively demanding computer memory as well as the calculation time 
if we used a model space consistent with that adopted in the HFB calculation. 
Accordingly, we need another factor $f_{pp}$
for the particle-particle channel. 
We determine this factor such that the spurious $K^{\pi}=0^{+}$ mode 
associated with the particle number fluctuation appears at zero energy
($v_{pp} \rightarrow f_{pp}\cdot v_{pp}$). 

\section{\label{check}Check of the calculation scheme}
\begin{figure}[t]
\begin{center}
\includegraphics[scale=0.48]{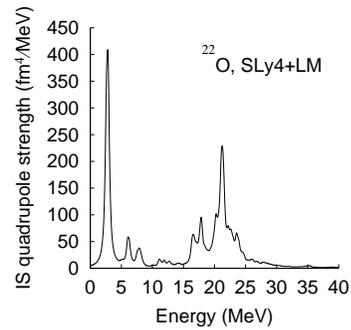}
\caption{Response function for the isoscalar quadrupole operator in $^{22}$O. 
The transition strengths are smeared by using a Lorentzian function with a width of $\Gamma=0.5$ MeV. 
The renormalization factors for the QRPA calculation are $f_{ph}=0.982, f_{pp}=1.18$.
The cutoff energy is 60 MeV. 
}
\label{22O_response}
\end{center}
\end{figure}

In this section, we compare our results with those of Ref.~\cite{kha02}. 
In this reference, the SLy4 interaction~\cite{cha98} for the mean field 
and the surface-type delta interaction with $\gamma=1.5$ and 
$V_{0}=-415.73$ MeV$\cdot$fm$^{3}$ for the pairing field were employed for the 
HFB calculation, and the quasiparticle energy was cut off at 50 MeV. 
Therefore, we adopt these parameters for the comparisons in this section.  
The differences between the present calculation and that in Ref.~\cite{kha02} are 
the mesh size, the boundary condition, the cutoff energy for the QRPA calculation, 
and the treatment of the spin-dependent interaction ($G_{0}$ and $G_{0}^{\prime}$) 
in Eq.~(\ref{eq:res_ph}). 
In the present calculation, the spin transition density is treated exactly.

\begin{figure}[b]
\begin{center}
\includegraphics[scale=0.65]{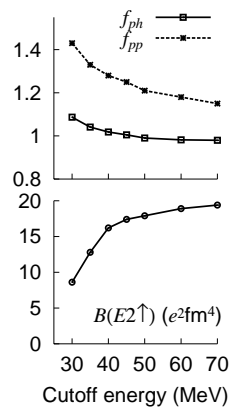}
\caption{Cutoff energy dependence of the renormalization factors 
and the $B(E2\uparrow)$ value for the first $2^{+}$ state in $^{22}$O.
}
\label{22O_dep}
\end{center}
\end{figure}

In Fig.~\ref{22O_response}, we show the isoscalar quadrupole response function
in $^{22}$O. The first $2^{+}$ state is located at 2.8 MeV with 
$B(E2\uparrow)=18.9$ $e^{2}$fm$^{4}$. 
The experimental values are $E(2^{+})_{\mathrm{exp}}=3.2$ MeV and 
$B(E2)_{\mathrm{exp}}=21\pm 8$ $e^{2}$fm$^{4}$~\cite{bel01,thi00,bec06}.
In Ref.~\cite{kha02}, the energy and the transition strength are 
$E(2_{1}^{+})=1.9$ MeV and $B(E2)=22$ $e^{2}$fm$^{4}$.  
The energy and the transition strength of the 
low-lying collective state is quite sensitive to the cutoff energy for 
the RPA calculation~\cite{bla77}. 
In Fig.~\ref{22O_dep}, the cutoff energy dependence of the renormalization factors 
and the $B(E2\uparrow)$ value for the $2_{1}^{+}$ state in $^{22}$O are shown. 
Even with the cutoff energy of 70 MeV, the transition strength for the low-lying 
state does not converge yet. 
In this case, the dimension of the QRPA matrix in Eq.~(\ref{eq:AB1}) is 11726 for 
the $K^{\pi}=0^{+}$ channel and the memory size is 13 GB, 
and the CPU time is about 70,000s per each iteration 
for determining the renormalization factor $f_{pp}$. 
If we could perform the QRPA calculation including all quasiparticle states obtained 
in the HFB calculation, the renormalization factor for the pairing channel $f_{pp}$ 
would be 1, because the p-p channel is treated self-consistently 
between the HFB and the QRPA calculations. 

The peak position of the giant resonance is located slightly higher 
than in Ref.\cite{kha02}. 
The non-collective two-quasiparticle states around 6 and 7 MeV are consistent 
between the two calculations. 
The energy-weighted sum ($1.867\times10^{4}$MeV$\cdot$fm$^{4}$)
overestimates by about 13.9\% the EWSR value ($1.638\times 10^{4}$MeV$\cdot$fm$^{4}$). 
The overshooting of the EWSR for the isoscalar quadrupole mode in the
LM approximation was pointed out in Ref~\cite{miz07}.

\begin{figure*}[t]
\begin{center}
\includegraphics[scale=0.8]{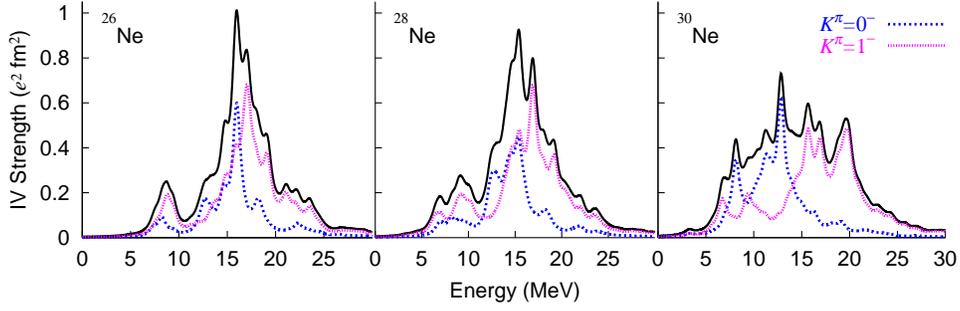}
\caption{Response functions for the isovector dipole operator in $^{26,28,30}$Ne. 
The dotted, dashed and solid lines correspond to the $K^{\pi}=0^{-}$, $K^{\pi}=1^{-}$ and 
total responses, respectively. 
For the $K^{\pi}=1^{-}$ response, the transition strengths for the $K^{\pi}=\pm1^{-}$ states 
are summed up. The transition strengths are smeared by using  $\Gamma=1$ MeV. 
The renormalization factors for the QRPA calculation 
are $f_{ph}=0.919, 0.880$ and 0.929 for $^{26,28,30}$Ne and $f_{pp}=1.225$ for all nuclei. 
}
\label{response}
\end{center}
\end{figure*}

\section{\label{results}Results and Discussion}
\begin{table}[b]
\caption{Ground state properties of $^{26,28,30}$Ne obtained by the
deformed HFB calculation with the SkM* interaction and the surface-type pairing interaction. 
Chemical potentials, deformations, average pairing gaps and 
root-mean-square radii for neutrons and protons are listed.
}
\label{GS}
\begin{center} 
\begin{tabular}{cccc} \hline \hline
 & $^{26}$Ne & $^{28}$Ne & $^{30}$Ne \\ \hline
$\lambda_{\nu}$ (MeV) & $-4.60$ & $-3.06$ & $-2.90$ \\
$\lambda_{\pi}$ (MeV) & $-14.8$ & $-17.0$ & $-19.9$ \\
$\beta_{2}^{\nu}$ & 0.08 & 0.12 & 0.32 \\
$\beta_{2}^{\pi}$ & 0.14 & 0.20  & 0.39 \\
$\langle \Delta_{\nu} \rangle$ (MeV) & 0.0 (0.70)  & 1.27 (1.24)  & 1.34 (1.30)  \\
$\langle \Delta_{\pi} \rangle$ (MeV) & 1.04 & 0.87 & 0.0 \\
$\sqrt{\langle r^{2} \rangle_{\nu}}$ (fm) & 3.20 & 3.35 & 3.53 \\
$\sqrt{\langle r^{2} \rangle_{\pi}}$ (fm) & 2.93 & 2.98 & 3.08 \\ \hline \hline
\end{tabular}
\end{center} 
\end{table}

We now discuss the properties of $^{26,28,30}$Ne nuclei calculated with the SkM* interaction. 
We summarize in Table~\ref{GS} the ground state properties of these Ne isotopes 
obtained by solving Eq.~(\ref{eq:HFB1}). 
The ground state is slightly deformed in $^{26}$Ne and $^{28}$Ne, 
and we obtain a well-deformed ground state for $^{30}$Ne. 
The values in parentheses are experimental pairing gaps extracted by 
the three-point mass difference formula~\cite{sat98} 
using the experimental binding energies taken from Ref.~\cite{aud95}.  
We define the deformation parameter $\beta_{2}$ and average pairing gap 
$\langle\Delta\rangle$~\cite{sau81,ben00,dug01,yam01} as
\begin{align}
\beta_{2}^{\tau}&=\dfrac{4\pi}{5}\dfrac{\int d\bold{r} \varrho^{\tau}(\bold{r})r^{2}Y_{20}(\hat{r})}
{\int d\bold{r} \varrho^{\tau}(\bold{r})r^{2}},\\
\langle\Delta_{\tau}\rangle&=-\dfrac{\int d\bold{r} \tilde{\varrho}^{\tau}(\bold{r})
\tilde{h}^{\tau}(\bold{r})}{\int d\bold{r} \tilde{\varrho}^{\tau}(\bold{r})},
\end{align} 
where $\tilde{\varrho}(\boldsymbol{r})$ is the pairing density.

Fig.~\ref{response} shows the response functions for the isovector dipole mode in 
neutron-rich Ne isotopes. 
The isovector dipole operator used in the present calculation is
\begin{equation}
\hat{F}_{1K}=e\dfrac{N}{A}\sum_{i}^{Z}r_{i}Y_{1K}(\hat{r}_{i})-
e\dfrac{Z}{A}\sum_{i}^{N}r_{i}Y_{1K}(\hat{r}_{i}),
\end{equation}
and the response functions are calculated as
\begin{equation}
S(E)=\sum_{i}\sum_{K} \dfrac{\Gamma/2}{\pi}\dfrac{|\langle i|\hat{F}_{1K}|0\rangle|^{2}}
{(E-\hbar \omega_{i})^{2}+\Gamma^{2}/4}.
\end{equation} 

\subsection{$^{26}$Ne}

We can clearly see a resonance structure at around the excitation energy of 8-9 MeV, 
together with the giant resonance at $15-20$ MeV. 
Because of the small deformation the $K$ splitting is small and smeared out. 

\begin{figure}[t]
\begin{center}
\includegraphics[scale=0.7]{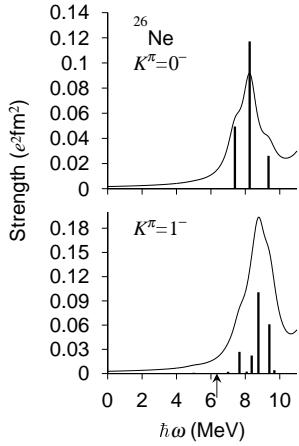}
\caption{Isovector dipole transition strengths in $^{26}$Ne 
for the $K^{\pi}=0^{-}$ (the upper) and $K^{\pi}=1^{-}$ (the lower) states.
Underlying discrete states are shown together with the smeared response functions. 
The arrow indicates the neutron emission threshold $E_{\mathrm{th}}=6.58$ MeV.
}
\label{strength}
\end{center}
\end{figure}

\begin{table}[b]
\caption{QRPA amplitudes for the $K^{\pi}=0^{-}$ state in $^{26}$Ne at 8.25 MeV.
This mode has the proton strength $B(E1)=2.98 \times10^{-2}~e^{2}$fm$^{2}$, 
the neutron strength $B(Q^{\nu}1)=2.89 \times10^{-2} e^{2}$fm$^{2}$, 
and the isovector strength $B(Q^{\mathrm{IV}}1)=1.17 \times10^{-1} e^{2}$fm$^{2}$, 
and the sum of backward-going amplitude $\sum|Y_{\alpha\beta}|^{2}=4.33\times 10^{-3}$. 
The single-(quasi)particle levels are labeled with 
the asymptotic quantum numbers $[Nn_{3}\Lambda]\Omega$. 
Only components with $X_{\alpha\beta}^{2}-Y_{\alpha\beta}^{2} > 0.001$ are listed. 
Two-quasiparticle excitation energies are given by $E_{\alpha}+E_{\beta}$ in MeV and 
two-quasiparticle transition matrix elements $Q_{10,\alpha\beta}$ in $e \cdot$fm.
In the row (i), the label 
$\nu 1/2^{-}$ denotes a non-resonant discretized continuum state of neutron 
$\Omega^{\pi}=1/2^{-}$ level.
}
\label{26Ne_0-}
\begin{center} 
\begin{tabular}{cccccc}
\hline \hline
 &  &  & $E_{\alpha}+E_{\beta}$ &  & 
$Q_{10,\alpha\beta}$  \\
 & $\alpha$ & $\beta$ & (MeV) & $X_{\alpha \beta}^{2}-Y_{\alpha\beta}^{2}$ & ($e\cdot$ fm) 
\\ \hline
(a) & $\nu[310]1/2$ & $\nu[211]1/2$ & 8.15 & 0.670 & $-0.309$ \\ 
(b) & $\nu[330]1/2$ & $\nu[220]1/2$ & 11.4 & 0.020 & $-0.397$ \\
(c) & $\nu[312]5/2$ & $\nu[202]5/2$ & 11.2 & 0.006 & $-0.239$ \\
(d) & $\nu[321]3/2$ & $\nu[211]3/2$ & 11.3 & 0.006 & 0.338 \\ 
(e) & $\nu[330]1/2$ & $\nu[211]1/2$ & 6.54 & 0.003 & $-0.118$ \\
(f) & $\nu[312]3/2$ & $\nu[211]3/2$ & 12.8 & 0.002 & $-0.014$ \\
(g) & $\nu[301]1/2$ & $\nu[211]1/2$ & 9.32 & 0.002 & $-0.117$ \\ 
(h) & $\nu[200]1/2$ & $\nu[101]1/2$ & 14.0 & 0.002 & $-0.241$ \\  
(i) & $\nu 1/2^{-}$ & $\nu[211]1/2$ & 12.6 & 0.002 & $-0.068$ \\
\hline
(j) & $\pi[220]1/2$ & $\pi[101]1/2$ & 7.96 & 0.265 & 0.0085 \\
(k) & $\pi[330]1/2$ & $\pi[220]1/2$ & 13.4 & 0.008 & $-0.329$ \\
(l) & $\pi[220]1/2$  & $\pi[110]1/2$ & 14.1 & 0.008 & $-0.346$ \\
\hline \hline
\end{tabular}
\end{center} 
\end{table}
\begin{table}[t]
\caption{Same as Table~\ref{26Ne_0-} but for the $K^{\pi}=1^{-}$ state in $^{26}$Ne at 8.76 MeV.
This mode has $B(E1)=1.65 \times10^{-2}~e^{2}$fm$^{2}$, 
$B(Q^{\nu}1)=3.58 \times10^{-2} e^{2}$fm$^{2}$, 
$B(Q^{\mathrm{IV}}1)=1.00 \times10^{-1} e^{2}$fm$^{2}$, 
and $\sum|Y_{\alpha\beta}|^{2}=2.93\times 10^{-3}$. 
}
\label{26Ne_1-}
\begin{center} 
\begin{tabular}{cccccc}
\hline \hline
 &  &  & $E_{\alpha}+E_{\beta}$ &  & 
$Q_{11,\alpha\beta}$  \\
 & $\alpha$ & $\beta$ & (MeV) & $X_{\alpha \beta}^{2}-Y_{\alpha\beta}^{2}$ & ($e\cdot$ fm) 
\\ \hline
(a) & $\nu[312]3/2$ & $\nu[211]1/2$ & 8.68 & 0.849 & 0.339 \\
(b) & $\nu[310]1/2$ & $\nu[211]1/2$ & 8.16 & 0.040 & $-0.131$ \\
(c) & $\nu[301]1/2$ & $\nu[211]1/2$ & 9.32 & 0.010 & 0.294 \\
(d) & $\nu[321]3/2$ & $\nu[220]1/2$ & 12.0 & 0.007 & 0.250 \\
(e) & $\nu[303]7/2$ & $\nu[202]5/2$ & 12.1 & 0.006 & 0.414 \\
(f) & $\nu[330]1/2$ & $\nu[220]1/2$ & 11.4 & 0.004 & $-0.127$ \\
(g) & $\nu[312]5/2$ & $\nu[211]3/2$ & 12.1 & 0.004 & 0.348 \\
(h) & $\nu[321]3/2$ & $\nu[202]5/2$ & 10.3 & 0.001 & $-0.010$ \\
(i) & $\nu[321]3/2$ & $\nu[202]5/2$ & 11.8 & 0.003 & $-0.214$ \\
(j) & $\nu[330]1/2$ & $\nu[211]1/2$ & 6.54 & 0.003 & $-0.081$ \\
(k) & $\nu[321]3/2$ & $\nu[211]1/2$ & 7.14 & 0.001 & 0.106 \\
 \hline
(l) & $\pi[220]1/2$ & $\pi[101]1/2$ & 7.96 & 0.037 & 0.0095 \\ 
(m) & $\pi[211]3/2$ & $\pi[101]1/2$ & 7.95 & 0.015 & $-0.011$ \\
(n) & $\pi[321]3/2$ & $\pi[220]1/2$ & 14.0 & 0.004 & 0.313 \\
(o) & $\pi[312]5/2$ & $\pi[211]3/2$ & 14.7 & 0.002 & $-0.338$ \\
(p) & $\pi[211]3/2$ & $\pi[110]1/2$ & 14.1 & 0.002 & 0.280 \\
(q) & $\pi[211]1/2$ & $\pi[101]1/2$ & 11.5 & 0.002 & $-0.256$ \\
\hline \hline
\end{tabular}
\end{center} 
\end{table}

In Fig.~\ref{strength}, we show the transition strengths in the low-energy region. 
The neutron emission threshold is 6.35 MeV, and 
the resonance which is composed of several discrete states appears just above the threshold. 
In contrast to the low-lying quadrupole state in $^{22}$O, the transition strengths 
for the dipole states in this region converge at the cutoff energy of about 40 MeV. 
We made a detailed analysis of the QRPA eigenmodes 
and show in Tables~\ref{26Ne_0-},\ref{26Ne_1-} the microscopic structures  
of the $K^{\pi}=0^{-}$ state at 8.25 MeV 
and the $K^{\pi}=1^{-}$ state at 8.76 MeV, 
which have the largest transition strength for each sector.
In the Tables, single-(quasi)particle states are labeled 
with the asymptotic quantum numbers 
$[Nn_{3}\Lambda]\Omega$ just for convenience.
It should be noted that 
the asymptotic quantum numbers are not good quantum numbers 
because the deformation is not so large. 

For the $K^{\pi}=0^{-}$ state at 8.25 MeV, 
the dominant component is the $\nu[211]1/2 \to \nu[310]1/2$ transition, corresponding 
to $\nu(2s^{-1}_{1/2} 2p_{3/2})$. 
Particle-hole excitations of (b), (c) and (d) correspond to 
$\nu(1d^{-1}_{5/2}1f_{7/2})$ excitation, 
which have 3.2\% contribution in total. 
The rows (e), (f), (g) and (h) correspond to $\nu(2s^{-1}_{1/2}1f_{7/2})$, 
$\nu(1d^{-1}_{5/2}2p_{3/2})$, $\nu(2s^{-1}_{1/2}2p_{1/2})$ 
and $\nu(1p^{-1}_{1/2}1d_{3/2})$ excitations, respectively. 
Two-quasiparticle proton excitations, furthermore, have an appreciable contribution; 
the rows (j), (k), and (l) correspond to 
the hole-hole like $\pi (1p_{1/2} \otimes 1d_{5/2})$ excitation, 
particle-hole like $\pi (1d^{-1}_{5/2}1f_{7/2})$, and 
$\pi (1d^{-1}_{5/2}1p_{3/2})$ excitations, respectively.
The $K^{\pi}=1^{-}$ state has a similar structure to the $K^{\pi}=0^{-}$ state.
The main component is $\nu(2s^{-1}_{1/2} 2p_{3/2})$, which corresponds to 
(a) and (b) in Table~\ref{26Ne_1-}. 
With a small contribution, many other neutron particle-hole and 
proton two-quasiparticle excitations build the excitation mode at 8.76 MeV. 
The resonance is also composed of the $K^{\pi}=1^{-}$ mode appearing at 9.40 MeV. 
This mode is dominantly (97.6\%) generated by the $\nu[211]1/2 \to \nu[301]1/2$ transition 
corresponding to the $\nu(2s_{1/2}^{-1}2p_{1/2})$ transition. 

Preliminary calculations of deformed QRPA using the Gogny interaction~\cite{per07}, 
and the relativistic deformed QRPA~\cite{pen07} show that the low-lying dipole 
state is dominantly constructed by the $\nu(2s^{-1}_{1/2}2p_{3/2})$ configuration. 

\begin{figure}[t]
\begin{center}
\begin{tabular}{cc}
\includegraphics[scale=0.4]{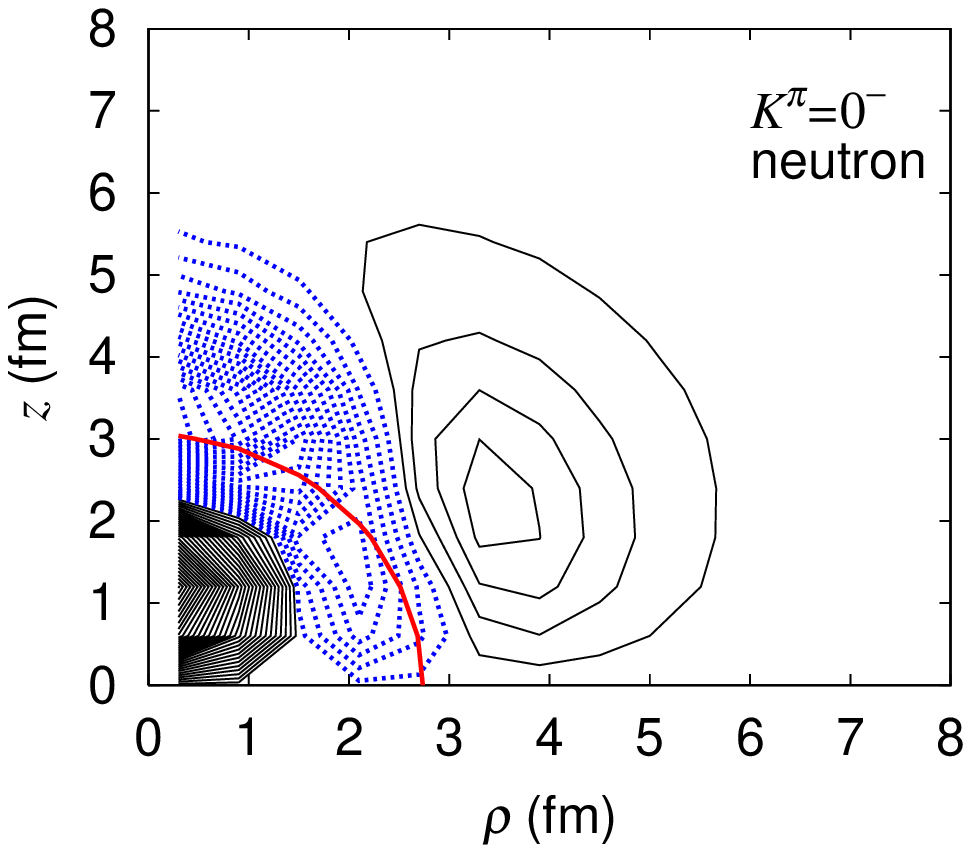}
\includegraphics[scale=0.4]{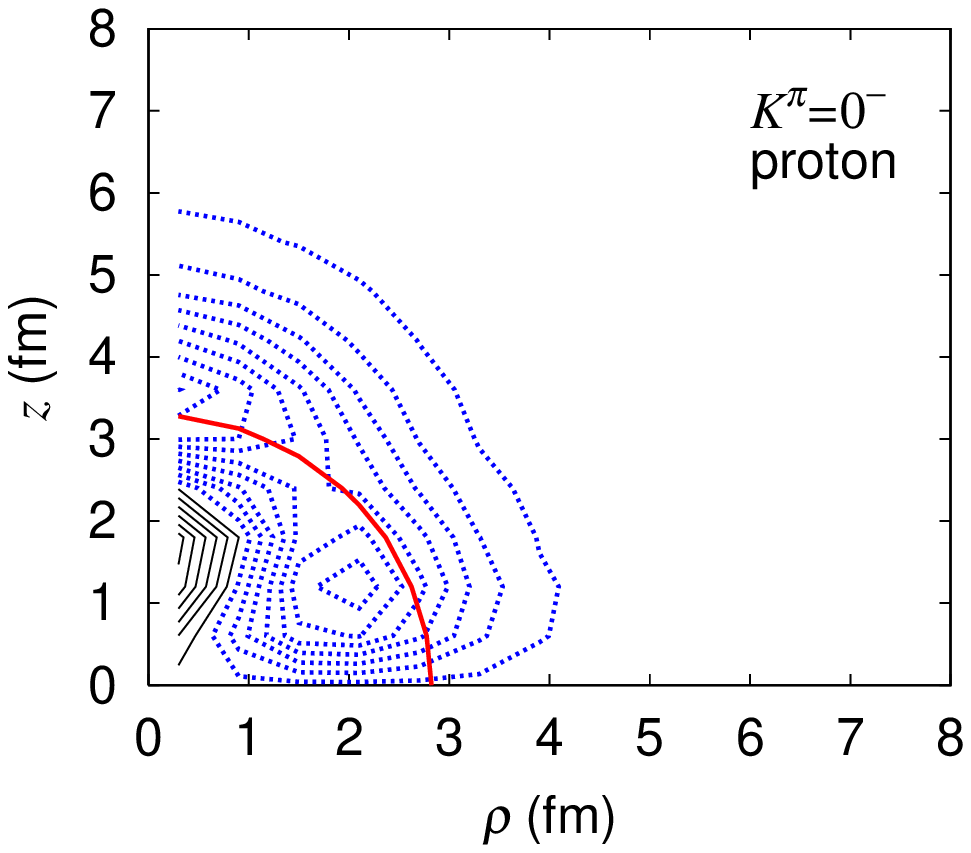} \\
\includegraphics[scale=0.4]{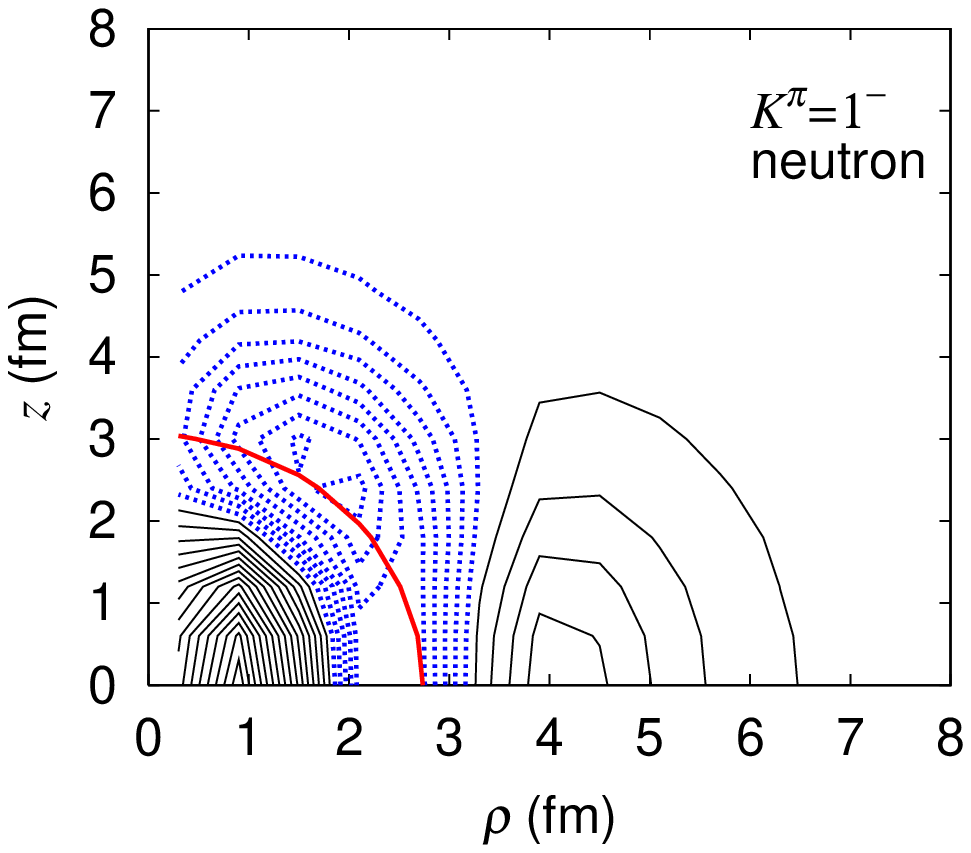}
\includegraphics[scale=0.4]{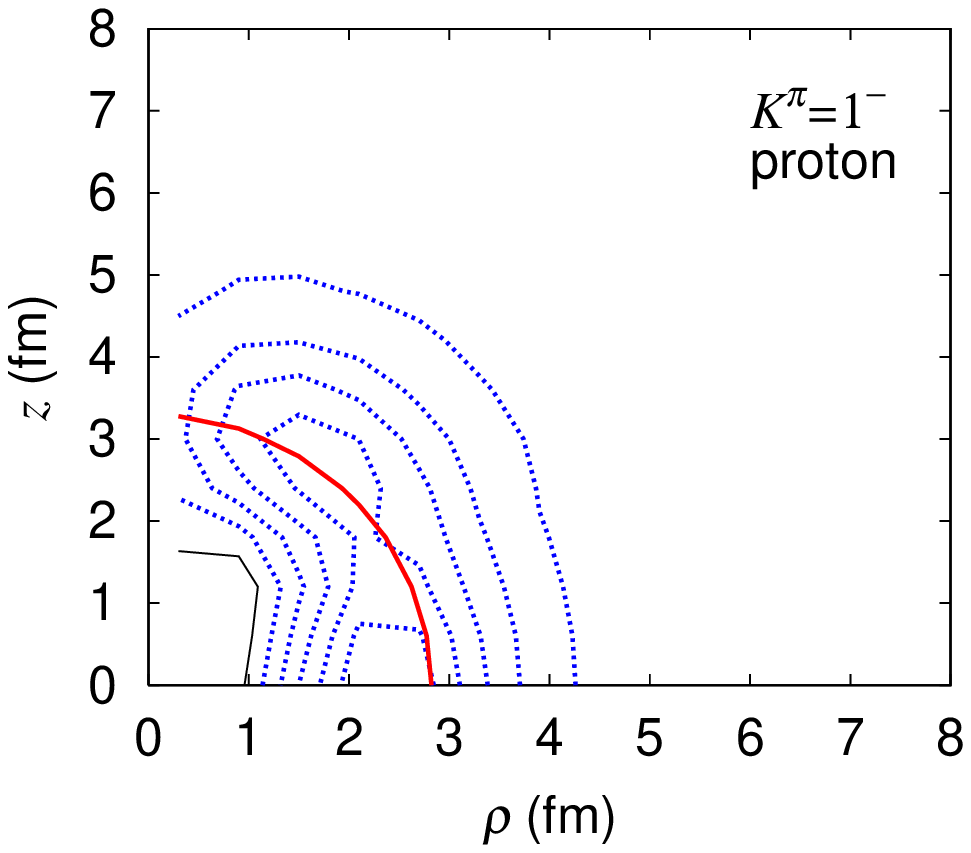}
\end{tabular}
\caption{Transition densities in $^{26}$Ne 
for the $K^{\pi}=0^{-}$ state at 8.25 MeV (upper panels), 
and for the $K^{\pi}=1^{-}$ state at 8.76 MeV (lower panels). 
Solid and dotted lines indicate positive 
and negative transition densities, and 
the contour lines are plotted at intervals of $3 \times 10^{-4}$ fm$^{-3}$. 
The thick solid lines indicate the neutron and proton half density, 
0.058 fm$^{-3}$ and 0.036 fm$^{-3}$, respectively.
}
\label{trans_density}
\end{center}
\end{figure}

\begin{figure}[t]
\begin{center}
\includegraphics[scale=0.4]{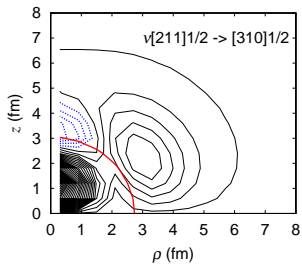}
\caption{Same as Fig.~\ref{trans_density}, for the unperturbed transition 
density of $\nu[211]1/2 \to [310]1/2$ excitation.}
\label{26Ne_wf}
\end{center}
\end{figure}

In Fig.~\ref{trans_density}, we show transition densities for 
the $K^{\pi}=0^{-}$ state at 8.25 MeV, and for the $K^{\pi}=1^{-}$ state at 8.76 MeV. 
These transition densities are quite different from the classical picture 
of the isovector giant resonances. 
They have an isoscalar character in the surface region of the nucleus. 
On the other hand, outside of the nucleus, neutrons have an oscillation and 
the neutron excitation is dominant. Furthermore, 
the neutron excitations take place in the low-density region around 6 fm, 
namely, this mode has a unique picture of vibration of the neutron skin. 
The spatially extended structure of the $\nu[211]1/2$ state is responsible for 
this tail of the neutron transition density. 
In order to clearly see the spatial structure of $\nu[211]1/2$, 
the unperturbed transition density of $\nu[211]1/2 \to [310]1/2$ is shown in 
Fig.~\ref{26Ne_wf}. 
It is clear that the wave function extends far outside of the nucleus, 
and the extension is larger than in the dipole state of Fig.~\ref{trans_density}. 
Furthermore, the structure around the surface is different between the unperturbed 
one and that obtained in the QRPA. 
These differences are generated by the QRPA correlations; the low-lying dipole 
state possesses a collective nature, which is small but finite. 

\begin{figure}[t]
\begin{center}
\includegraphics[scale=0.53]{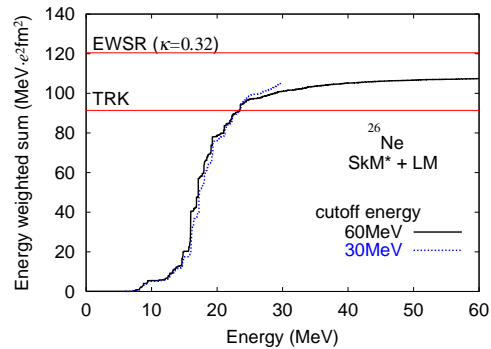}
\caption{Energy weighted sum of the isovector dipole strength function. 
The solid and the dotted lines are calculations with the energy cutoff at 
60 and 30 MeV. 
The horizontal lines show the classical TRK, and the RPA sum rule 
including the enhancement factor, $m_{1}=m_{1}^{\mathrm{cl}}(1+\kappa)$. 
Here $\kappa=0.32$ for the SkM* interaction in $^{26}$Ne.}
\label{EWSR}
\end{center}
\end{figure}

Fig.~\ref{EWSR} shows the energy weighted sum of the isovector dipole strength function 
together with the sum rule values represented by the horizontal lines. 
The calculated sum satisfies 89.2\% of the EWSR value 
including the enhancement factor $\kappa$; 
$m_{1}=m_{1}^{\mathrm{cl}}(1+\kappa)$~\cite{ter06}. 
The enhancement factor comes from the momentum dependence of 
the Skyrme density functionals. 
The effect of the explicit treatment of the momentum dependence for the EWSR 
was discussed in the discretized-continuum QRPA~\cite{yam02} 
and the continuum QRPA~\cite{miz07} for the spherical systems. 
In the present calculation, 
we treat the momentum dependence in the LM approximation. 
Therefore, discrepancy between the calculation and the EWSR value 
comes from this treatment of the momentum dependence. 
This point remains to be improved, and it is discussed in Ref.~\cite{yos08}.
 
In the present calculation, 
the energy-weighted sum up to 10 MeV is 5.51 MeV$\cdot e^{2}$fm$^{2}$, 
corresponding to 6.0\% of the TRK sum-rule value, 
4.6\% of the EWSR including the enhancement factor 
and 5.1\% of the calculated sum. 
These values are consistent with the experiment~\cite{gib07}. 
In Fig.~\ref{EWSR}, we also show the energy-weighted sum calculated with the energy 
cutoff at 30 MeV (dotted line). 
In the giant resonance region, two calculations give different results, while 
they are almost identical in the low-energy region. 
This is because the collectivity of the low-lying resonance is small, 
and consequently the transition strength is not very sensitive to the cutoff energy.

Before going to the neighboring nuclei, it should be noted that 
we obtain the collective octupole state at about 5.2 MeV, below the neutron threshold, 
with $B(E3\uparrow)=2458 e^{2}$fm$^{6}$, 
which corresponds to about 61 in Weisskopf units and 
the isoscalar transition strength is $2.60\times10^{4}$fm$^{6}$. 
The lowest $K^{\pi}=0^{-}$ state is located at 5.03 MeV and the sum of the backward-going 
amplitudes is 0.099. 
This state is generated by $\nu[211]1/2\to[330]1/2 (53.2\%)$, 
$\nu[202]5/2\to[312]5/2 (6.8\%)$, $\nu[211]1/2\to[310]1/2 (6.5\%)$, 
$\nu[220]1/2\to[330]1/2 (3.7\%)$, and 
$\pi[101]2/1\otimes[220]1/2 (14.8\%)$, $\pi[101]3/2\otimes[211]3/2 (2.2\%)$, 
$\pi[220]1/2\otimes[330]1/2 (2.1\%)$, $\pi[101]1/2\otimes[211]1/2 (1.0\%)$.  

\subsection{28Ne and 30Ne}
\begin{figure}[t]
\begin{center}
\includegraphics[scale=0.7]{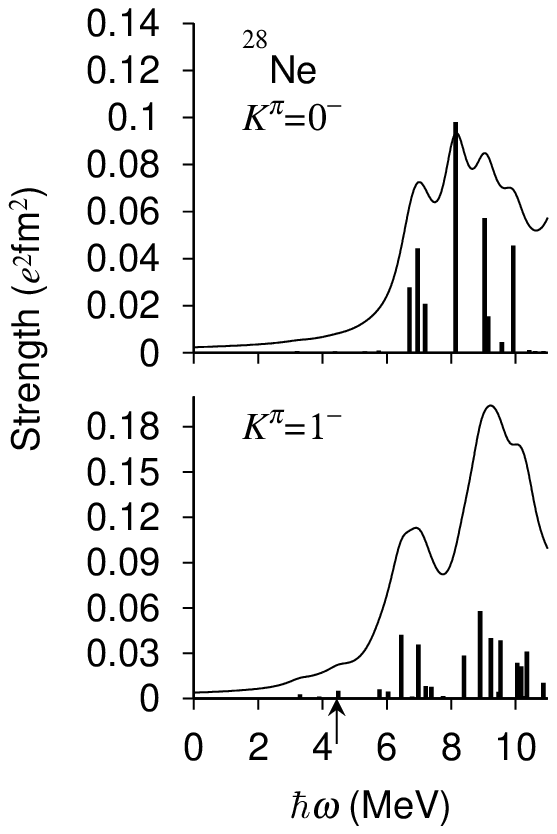}
\includegraphics[scale=0.7]{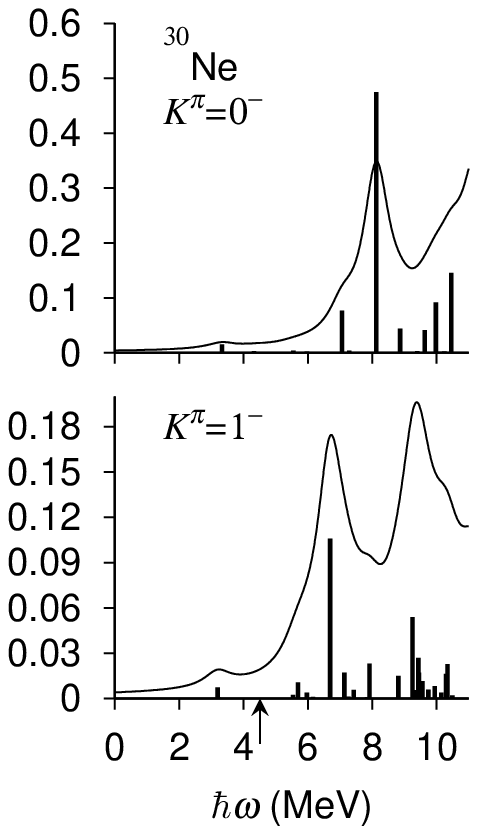}
\caption{Same as Fig.~\ref{strength} but for $^{28}$Ne and $^{30}$Ne. 
The arrows indicate the neutron emission threshold $E_{\mathrm{th}}=4.44$ and 4.51 MeV.
}
\label{28Ne_strength}
\end{center}
\end{figure}

The central panel in Fig.~\ref{response} shows the response function in $^{28}$Ne. 
In the low-energy region, we can see a two-bump structure at around 7 and 8 MeV. 
Because the deformation is small as in $^{26}$Ne, 
we cannot see a splitting of the giant resonance. 
In Fig.~\ref{28Ne_strength}, we show the low-energy part of the 
strength functions. 
In the $K^{\pi}=0^{-}$ states, there is a prominent peak at 8.1 MeV with 
a strength of 0.098 $e^{2}$fm$^{2}$. 
The strength distribution is fragmented for the $K^{\pi}=1^{-}$ mode, but 
correspondingly, we can see an eigenmode at 8.9 MeV 
with the largest transition strength of 0.058 $e^{2}$fm$^{2}$.

We show in Table~\ref{28Ne_0-} the QRPA amplitude for 
the $K^{\pi}=0^{-}$ state at 8.14 MeV in $^{28}$Ne. 
The main component is the neutron two-quasiparticle excitation of 
$\nu([310]1/2 \otimes [211]1/2)$ corresponding to $\nu(2s^{-1}_{1/2}2p_{3/2})$.
Two quasiparticle excitations of (b) and (c) in Table~\ref{28Ne_0-} correspond to 
$\nu(1d^{-1}_{5/2}1f_{7/2})$, and 
(d): $\nu(2s^{-1}_{1/2}1f_{7/2})$, (e) and (f): $\nu(1d^{-1}_{3/2}2p_{3/2})$, 
(g): $\nu(1d^{-1}_{3/2}1f_{7/2})$, (h) and (i): $\nu(1d^{-1}_{5/2}2p_{3/2})$, 
and (j): $\nu(1d^{-1}_{5/2}2p_{1/2})$ excitations, respectively. 
The proton excitation of $\pi(1p_{1/2}\otimes 1d_{5/2})$ has an appreciable 
contribution as in $^{26}$Ne.  

\begin{table}[t]
\caption{Same as Table~\ref{26Ne_0-} 
but for the $K^{\pi}=0^{-}$ state at 8.14 MeV in $^{28}$Ne.
This mode has $B(E1)=2.62 \times10^{-2}~e^{2}$fm$^{2}$, 
$B(Q^{\nu}1)=2.29 \times10^{-2} e^{2}$fm$^{2}$, 
$B(Q^{\mathrm{IV}}1)=9.80 \times10^{-2} e^{2}$fm$^{2}$, 
and $\sum|Y_{\alpha\beta}|^{2}=9.77\times 10^{-3}$. 
In the rows (k), (l), (m), (n) and (t), the labels 
$\nu 1/2^{-}$, $\nu 1/2^{+}$ and $\pi 1/2^{+}$ 
denote non-resonant discretized continuum states of neutron 
$\Omega^{\pi}=1/2^{-}$ and $1/2^{+}$ levels and proton $1/2^{+}$ level.
}
\label{28Ne_0-}
\begin{center} 
\begin{tabular}{cccccc}
\hline \hline
 &  &  & $E_{\alpha}+E_{\beta}$ &  & $Q_{10,\alpha\beta}$  \\
 & $\alpha$ & $\beta$ & (MeV) & $X_{\alpha \beta}^{2}-Y_{\alpha\beta}^{2}$ & ($e\cdot$ fm) 
\\ \hline
(a) & $\nu[310]1/2$ & $\nu[211]1/2$ & 8.27 & 0.569 & $-0.303$ \\
(b) & $\nu[330]1/2$ & $\nu[220]1/2$ & 11.2 & 0.055 & $-0.373$ \\
(c) & $\nu[321]3/2$ & $\nu[211]3/2$ & 10.9 & 0.006 & 0.323 \\ 
(d) & $\nu[330]1/2$ & $\nu[211]1/2$ & 6.20 & 0.036 & $-0.096$ \\
(e) & $\nu[312]3/2$ & $\nu[202]3/2$ & 6.82 & 0.004 & $-0.039$ \\
(f) & $\nu[310]1/2$ & $\nu[200]1/2$ & 5.81 & 0.003 & 0.026 \\
(g) & $\nu[330]1/2$ & $\nu[200]1/2$ & 3.74 & 0.004 & $-0.006$ \\   
(h) & $\nu[321]3/2$ & $\nu[211]3/2$ & 12.9 & 0.002 & $-0.014$ \\ 
(i) & $\nu[310]1/2$ & $\nu[220]1/2$ & 13.3 & 0.001 & $0.0004$ \\     
(j) & $\nu[301]1/2$ & $\nu[220]1/2$ & 14.5 & 0.001 & $-0.022$ \\
(k) & $\nu 1/2^{-}$ & $\nu[200]1/2$ & 10.1 & 0.009 & 0.203 \\
(l) & $\nu 1/2^{-}$ & $\nu[202]3/2$ & 10.7 & 0.002 & 0.099 \\
(m) & $\nu 1/2^{-}$ & $\nu[211]1/2$ & 12.6 & 0.004 & $-0.071$ \\
(n) & $\nu 1/2^{+}$ & $\nu[330]1/2$ & 14.2 & 0.003 & $-0.114$ \\ 
\hline
(o) & $\pi[220]1/2$ & $\pi[101]1/2$ & 7.63 & 0.238 & 0.018 \\
(p) & $\pi[330]1/2$ & $\pi[220]1/2$ & 12.8 & 0.031 & $-0.463$ \\
(q) & $\pi[220]1/2$ & $\pi[110]1/2$ & 13.7 & 0.007 & $-0.441$ \\
(r) & $\pi[211]3/2$ & $\pi[101]3/2$ & 12.3 & 0.004 & $-0.310$ \\
(s) & $\pi[211]1/2$ & $\pi[101]1/2$ & 11.6 & 0.002 & 0.288 \\
(t) & $\pi 1/2^{-}$ & $\pi[220]1/2$ & 19.3 & 0.001 & 0.004 \\
\hline \hline
\end{tabular}
\end{center} 
\end{table}

The lower energy resonance at around 7 MeV is described by three eigenstates 
as shown in Fig.~\ref{28Ne_strength}. 
The lower state at 6.70 MeV, which has an isovector strength of 
0.028 $e^{2}$fm$^{2}$, is mainly generated by $\nu(1d^{-2}_{3/2}2p_{3/2})$ (87.5\%), 
$\nu(2s^{-1}_{1/2}1f_{7/2})$ (5.7\%) and $\nu(1d^{-1}_{3/2}2p_{1/2})$ (2.4\%). 
The state at 6.96 MeV with $B(Q^{\mathrm{IV}}1)=0.044$ $e^{2}$fm$^{2}$ is 
almost a single p-h excitation of $\nu(1d^{-1}_{3/2}2p_{1/2})$ (90.2\%),
and the state at 7.19 MeV with 0.021 $e^{2}$fm$^{2}$ is generated dominantly 
by proton h-h like excitation of $\pi(1p_{1/2}\otimes 1d_{5/2})$ (57.8\%), 
together with $\nu(2s^{-1}_{1/2}1f_{7/2})$ (18.6\%), 
$\nu(1d^{-1}_{3/2}2p_{3/2})$ (6.4\%), $\nu(1d^{-1}_{5/2}1f_{7/2})$ (5.2\%), 
$\nu(2s^{-1}_{1/2}2p_{3/2})$ (4.5\%) and $\nu(1d^{-1}_{3/2}2p_{1/2})$ (1.8\%). 

Therefore, the higher-energy resonance at 8 MeV has a similar structure to 
that in $^{26}$Ne; $\nu(2s^{-1}_{1/2}2p_{3/2})$ and $\pi(1p_{1/2}\otimes 1d_{5/2})$ 
excitations are dominant, and the lower-energy resonance is generated by 
different eigenmodes. 

The $K^{\pi}=1^{-}$ state at 8.9 MeV in $^{28}$Ne has a similar structure 
to that in $^{26}$Ne and $K^{\pi}=0^{-}$ state in $^{28}$Ne, 
corresponding mainly to the neutron two-quasiparticle excitation of $\nu(2s^{-1}_{1/2}2p_{3/2})$, 
with 64.0\% contribution. 
In addition to this neutron p-h like excitation, the following excitations have 
an appreciable contribution; 
$\nu(1d^{-1}_{5/2}1f_{7/2})$ (13.9\%), 
$\nu(2s^{-1}_{1/2}1f_{7/2})$ (2.8\%), 
$\pi(1p^{-1}_{1/2}1d_{5/2})$ (7.9\%) and 
$\pi(1p_{3/2}\otimes 1d_{5/2})$ (1.4\%).

\begin{figure}[t]
\begin{center}
\begin{tabular}{cc}
\includegraphics[scale=0.42]{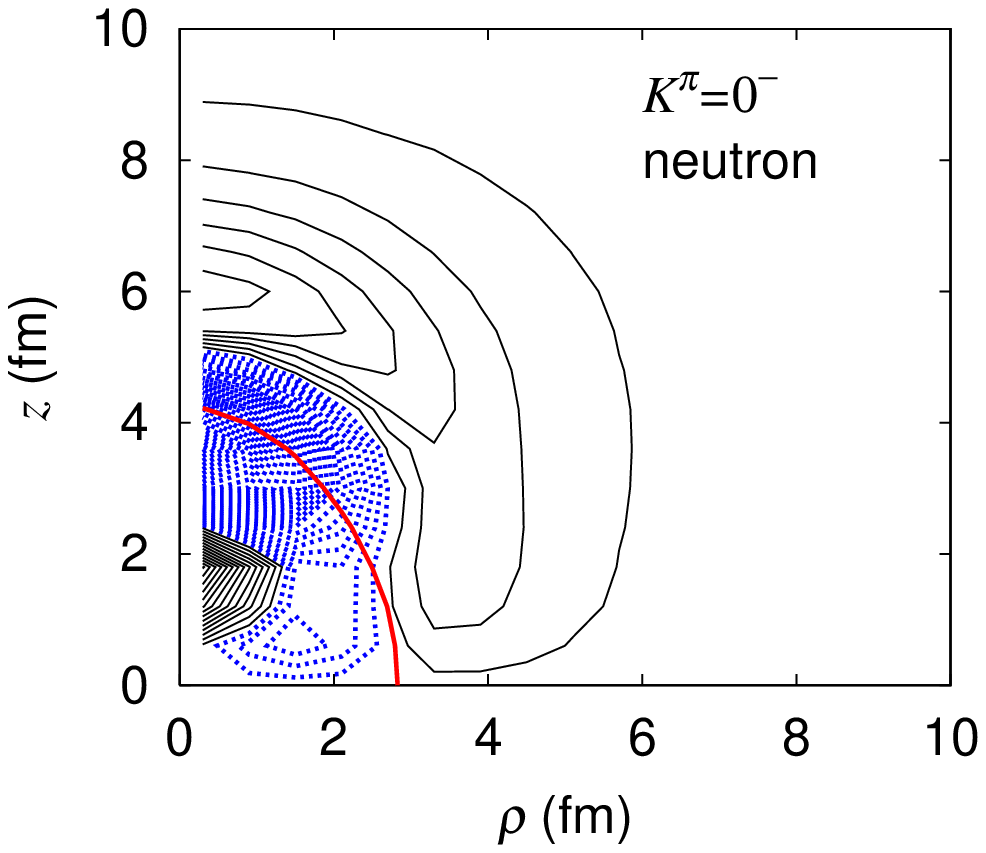}
\includegraphics[scale=0.42]{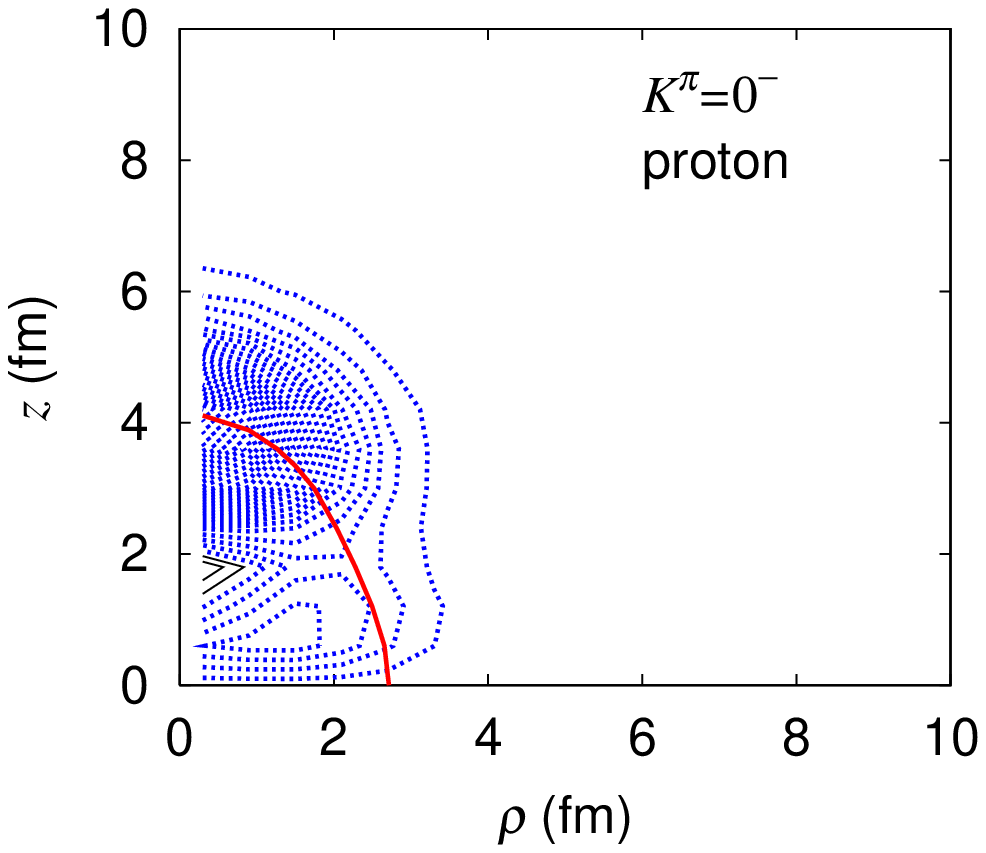} \\
\includegraphics[scale=0.42]{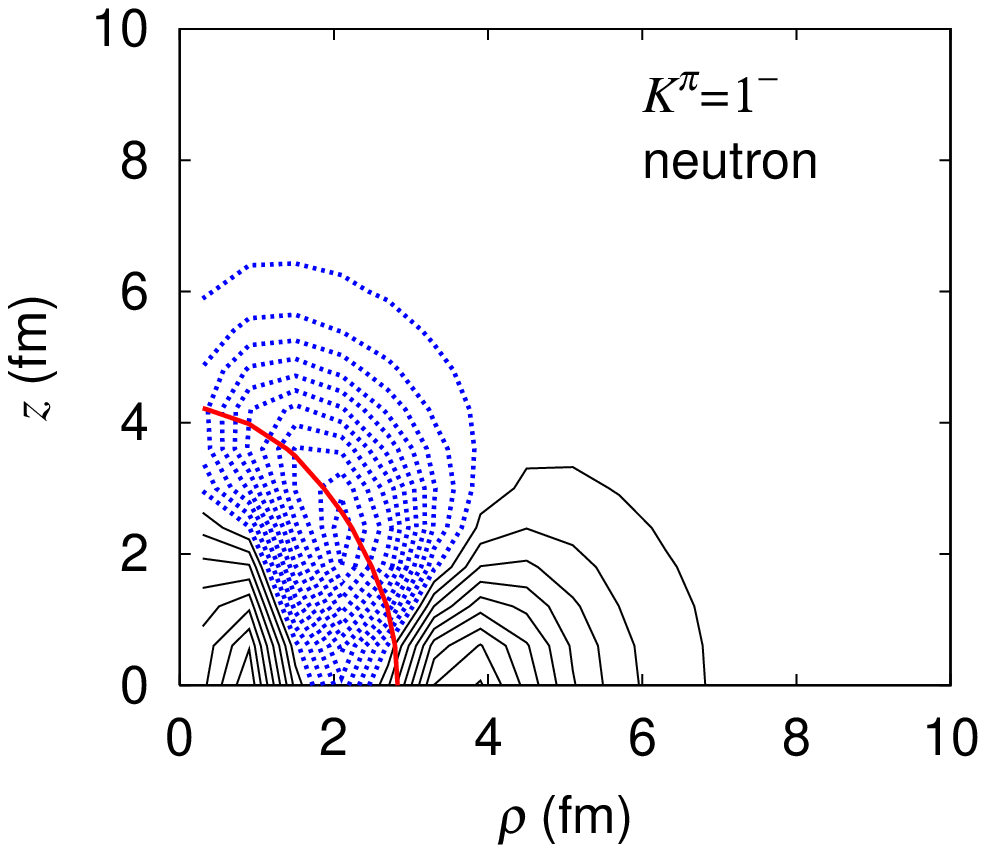}
\includegraphics[scale=0.42]{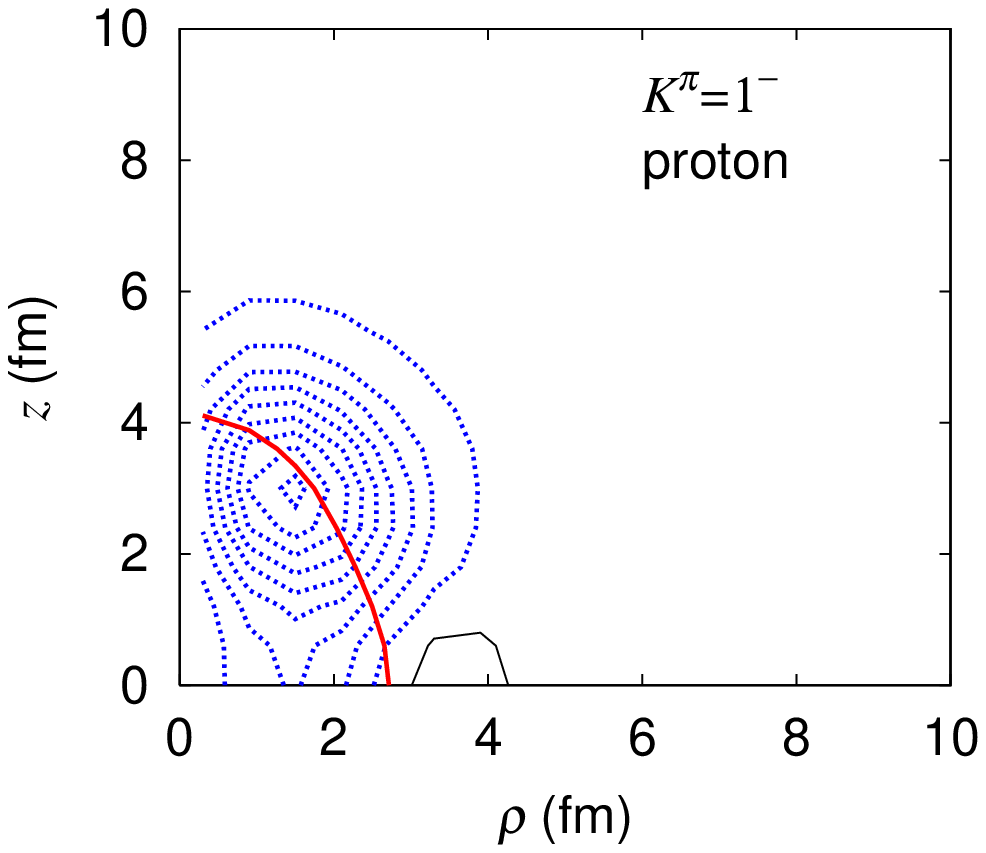}
\end{tabular}
\caption{Same as Fig.~\ref{trans_density} but for the $K^{\pi}=0^{-}$ 
state at 8.1 MeV and for the $K^{\pi}=1^{-}$ state at 6.7 MeV in $^{30}$Ne.
}
\label{30Ne_trans_density}
\end{center}
\end{figure}

Finally, we discuss the dipole state in $^{30}$Ne. 
Compared to the response functions in $^{26}$Ne and $^{28}$Ne, 
that for $^{30}$Ne is quite different, 
because this nucleus is well deformed as shown in Table~\ref{GS}. 
The giant resonance is split into $K^{\pi}=0^{-}$ and $1^{-}$ mode, 
and the split giant resonance has an overlap with the 
low-lying resonance below 10 MeV. 
In the right panel of Fig.~\ref{28Ne_strength}, 
we show the strength distribution below 10 MeV in $^{30}$Ne.
For the $K^{\pi}=0^{-}$ mode, we can see a prominent peak at 8.1 MeV 
possessing a large isovector $E1$ strength of 0.48 $e^{2}$fm$^{2}$. 
This state is mainly generated by $\nu[211]1/2 \to [310]1/2$ (38.4\%) and 
the neutron excitation from $[330]1/2$ to the non-resonance continuum state (37.9\%), 
together with the proton excitation of $\pi[330]1/2 \to [220]1/2$ (6.1\%). 

In Fig.~\ref{30Ne_trans_density}, the transition density for 
the $K^{\pi}=0^{-}$ state is shown. 
The transition density of protons are quite similar to that in 
Fig.~\ref{trans_density}. For the neutrons, 
we can easily see the effect of mixing 
of the excitation into the continuum state; 
the transition density has large spatial extension. 
Furthermore, comparing to Fig.~\ref{26Ne_wf}, this $K^{\pi}=0^{-}$ state 
still possesses a structure similar to the low-lying dipole state in $^{26}$Ne.

\begin{table}[t]
\caption{Same as Table~\ref{26Ne_0-} but for the $K^{\pi}=1^{-}$ state at 6.69 MeV 
in $^{30}$Ne, and only components with $X_{\alpha\beta}^{2}-Y_{\alpha\beta}^{2} > 0.01$ 
are listed. 
This mode has $B(E1)=1.58 \times10^{-2}~e^{2}$fm$^{2}$, 
$B(Q^{\nu}1)=3.98 \times10^{-2} e^{2}$fm$^{2}$, 
$B(Q^{\mathrm{IV}}1)=1.06 \times10^{-1} e^{2}$fm$^{2}$, 
and $\sum|Y_{\alpha\beta}|^{2}=8.82\times 10^{-3}$. 
}
\label{30Ne_1-}
\begin{center} 
\begin{tabular}{cccccc}
\hline \hline
 &  &  & $E_{\alpha}+E_{\beta}$ &  & 
$Q_{11,\alpha\beta}$  \\
 & $\alpha$ & $\beta$ & (MeV) & $X_{\alpha \beta}^{2}-Y_{\alpha\beta}^{2}$ & ($e\cdot$ fm) 
\\ \hline
(a) & $\nu[312]3/2$ & $\nu[200]1/2$ & 6.87 & 0.676 & 0.207 \\
(b) & $\nu[310]1/2$ & $\nu[200]1/2$ & 7.09 & 0.089 & 0.141 \\
(c) & $\nu[321]3/2$ & $\nu[202]5/2$ & 7.52 & 0.043 & $-0.009$ \\
(d) & $\nu[321]3/2$ & $\nu[211]1/2$ & 5.72 & 0.038 & 0.026 \\
(e) & $\nu[312]5/2$ & $\nu[202]3/2$ & 6.16 & 0.025 & 0.003 \\
(f) & $\nu[310]1/2$ & $\nu[202]3/2$ & 6.13 & 0.025 & 0.015 \\
(g) & $\nu[330]1/2$ & $\nu[211]1/2$ & 5.73 & 0.011 & 0.040 \\
(h) & $\nu1/2^{-}$ & $\nu[202]3/2$ & 7.57 & 0.019 & 0.066 \\
\hline \hline
\end{tabular}
\end{center} 
\end{table}

For the $K^{\pi}=1^{-}$ state, we can see a prominent peak at 6.69 MeV 
possessing an isovector strength of 0.11 $e^{2}$fm$^{2}$. 
This state has a different structure to the dipole states discussed above. 
In Table~\ref{30Ne_1-}, we show its microscopic structure. 
This state has a collective nature in a sense that a number of 
two-quasiparticle excitations have an appreciable contribution; 
in the present case, eight of the neutron excitations 
have a contribution larger than 1\%. 
In the lower panel of Fig.~\ref{30Ne_trans_density}, 
the transition density of this state is shown. 
This mode has also a characteristic feature that the neutron and proton 
contribution have an isoscalar nature around the surface region, and the neutron excitation 
is dominant outside of the nucleus.

It is difficult to link directly with the low-lying dipole states in $^{26}$Ne 
or $^{28}$Ne, because the deformations are quite different in $^{30}$Ne and in 
the other two nuclei. 
The main component of this $K^{\pi}=1^{-}$ state is (a):$\nu[200]1/2 \to [312]3/2$ and 
(b):$\nu[200]1/2 \to [310]1/2$. These p-h excitations are 
$\nu(1d^{-1}_{3/2}2p_{3/2})$ in the spherical limit. 
In this sense, the lower-energy resonance in $^{28}$Ne 
is connected to this collective $K^{\pi}=1^{-}$ state in $^{30}$Ne.

\section{\label{summary}Summary}
We have investigated a new framework of the deformed QRPA based on the 
Skyrme density functionals and the Landau-Migdal approximation. 
With this method, we have made a detailed analysis of the low-lying 
dipole states in neutron-rich $^{26,28,30}$Ne. 
In these nuclei, we obtain the excitation mode at $8-8.5$ MeV. 
The low-lying resonance is composed of several QRPA eigenmodes. 
In $^{26}$Ne, not only the $\nu(2s_{1/2}^{-1}2p_{3/2})$ transition 
but also the $\nu(2s_{1/2}^{-1}2p_{1/2})$ transition contribute to 
generating the resonance. 
In $^{28}$Ne and $^{30}$Ne, the $\nu[211]1/2 \to [310]1/2$ excitation still 
plays a major role. 
Each eigenmode is, however, not purely a single particle-hole excitation, 
it has a small contribution of the other neutron excitations and proton 
excitations as well. 

We have clearly shown the spatially extended structure of the 
$\nu[211]1/2$ ($2s_{1/2}$) state, and that it is responsible for the 
oscillation of transition density of neutrons outside of the nucleus. 
In the well deformed nucleus $^{30}$Ne, 
the deformation splitting of the giant resonance is large and 
the low-lying resonance overlaps with the giant resonance. 
For the $K^{\pi}=1^{-}$ state, we furthermore obtain a collective 
dipole mode at 6.7 MeV.

\begin{acknowledgments} 
The authors thank D.~Beaumel, L.~G.~Cao and members of the Groupe Th\'eorie 
in IPN Orsay for useful discussions and comments. 
One of the authors (K.Y) is supported by Research Fellowships of the 
Japan Society for the Promotion of Science for Young Scientists.
This work was supported by the JSPS Core-to-Core Program
``International Research Network for Exotic Femto Systems".  
The numerical calculations were performed on the NEC SX-8 supercomputers 
at Yukawa Institute for Theoretical Physics, Kyoto University and 
at Research Center for Nuclear Physics, Osaka University.
\end{acknowledgments}


\begin{thebibliography}{99}
\bibitem{tan01}
I.~Tanihata (Ed), 
Nucl. Phys. {\bf A693}, Nos. 1, 2 (2001).

\bibitem{hor01}
H.~Horiuchi, T.~Otsuka, Y.~Suzuki (Eds.),
Prog. Theor. Phys. Suppl. No.142 (2001).

\bibitem{hag02a}
K.~Hagino, H.~Horiuchi, M.~Matsuo, I.~Tanihata (Eds.), 
Prog. Theor. Phys. Suppl. No.146 (2002).

\bibitem{neu07}
P.~von Neumann-Cosel and T.~Aumann (Eds.), 
Nucl. Phys. {\bf A788}, (2007). 

\bibitem{suz95}
T.~Suzuki {\it et al.}, Phys. Rev. Lett. {\bf 75}, 3241 (1995).

\bibitem{miz00}
S.~Mizutori, J.~Dobaczewski, G.~A.~Lalazissis, W.~Nazarewicz, 
and P-.G.~Reinhard, Phys. Rev. C {\bf 61}, 044326 (2000), and references therein.

\bibitem{ike88}
K.~Ikeda, Nucl. Phys. {\bf A538}, 355c (1992). 

\bibitem{sac93}
D.~Sackett {\it et al.}, Phys. Rev. C {\bf 48}, 118 (1993).

\bibitem{shi95}
S.~Shimoura {\it et al.}, Phys. Lett. {\bf B348}, 29 (1995).

\bibitem{zin97}
M.~Zinser {\it et al.}, Nucl. Phys. {\bf A619}, 151 (1997).

\bibitem{nak06}
T.~Nakamura {\it et al.}, Phys. Rev. Lett. {\bf 96}, 252502 (2006).

\bibitem{nak94}
T.~Nakamura {\it et al.}, Phys. Lett. {\bf B331}, 296 (1994).

\bibitem{pal03}
R.~Palit {\it et al.}, Phys. Rev. C {\bf 68}, 034318 (2003).

\bibitem{fuk04}
N.~Fukuda {\it et al.}, Phys. Rev. C {\bf 70}, 054506 (2004).

\bibitem{nak99}
T.~Nakamura {\it et al.}, Phys. Phys. Lett. {\bf 83}, 1112 (1999).

\bibitem{pra03}
U.~D.~Pramanik {\it et al.}, Phys. Lett. {\bf B551}, 63 (2003).

\bibitem{aum99}
T.~Aumann {\it et al.}, Phys. Rev. C {\bf 59}, 1252 (1999).

\bibitem{lei01}
A.~Leistenschneider {\it et al.}, Phys. Rev. Lett. {\bf 86}, 5442 (2001).

\bibitem{try03}
E.~Tryggestad {\it et al.}, Phys. Rev. C {\bf 67}, 064309 (2003).

\bibitem{adr05}
P.~Adrich {\it et al.}, Phys. Rev. Lett. {\bf 95}, 132501 (2005).

\bibitem{cat97}
F.~Catara, E.~G.~Lanza, M.~A.~Nagarajan, and A.~Vitturi,
Nucl. Phys. {\bf A624}, 449 (1997).

\bibitem{ham98}
I.~Hamamoto, H.~Sagawa, and X.~Z.~Zhang, Phys. Rev. C {\bf 57}, R1064 (1998).

\bibitem{ham99}
I.~Hamamoto, H.~Sagawa, and X.~Z.~Zhang, Nucl. Phys. {\bf A648}, 203 (1999).

\bibitem{gor02}
S.~Goriely and E.~Khan, Nucl. Phys. {\bf A706}, 217 (2002).

\bibitem{mat02}
M.~Matsuo, Prog. Theor. Phys. Suppl. {\bf 146}, 110 (2002).

\bibitem{ter06}
J.~Terasaki, J.~Engel, Phys. Rev. C {\bf 74}, 044301 (2006).

\bibitem{col01}
G.~Col\`o, P.~F.~Bortignon, Nucl. Phys. {\bf A696}, 427 (2001).

\bibitem{sar04}
D.~Sarchi, P.~F.~Bortignon and G.~Col\`o, Phys. Lett. {\bf B601}, 27 (2004).

\bibitem{vre01a}
D.~Vretenar, N.~Paar, P.~Ring, G.~A.~Lalazissis,
Phys. Rev. C {\bf 63}, 047301 (2001). 

\bibitem{vre01b}
D.~Vretenar, N.~Paar, P.~Ring, G.~A.~Lalazissis, 
Nucl. Phys. {\bf A692}, 496 (2001).

\bibitem{paa05}
N.~Paar, T.~Nik\v si\'c, D.~Vretenar, P.~Ring, 
Phys. Lett. {\bf B606}, 288 (2005).

\bibitem{cao05}
L.~G.~Cao, Z.~Y.~Ma, 
Phys. Rev. C {\bf 71}, 034305 (2005).

\bibitem{liv07}
E.~Litivinova,~P.~Ring, and D.~Vretenar, Phys. Lett. {\bf B647}, 111 (2007).


\bibitem{gib07}
J.~Gibelin {\it et al.}, Nucl. Phys. {\bf A788}, 153c (2007).

\bibitem{kha05}
E.~Khan, N.~Sandulescu, and N.~Van Giai, Phys. Rev. C {\bf 71} (2005) R042801.

\bibitem{yos06}
K.~Yoshida, M.~Yamagami, K.~Matsuyanagi, 
Nucl. Phys. {\bf A779}, 99 (2006).

\bibitem{bul80}
A.~Bulgac, Preprint No. FT-194-1980, 
Institute of Atomic Physics, Bucharest, 1980. 
[arXiv:nucl-th/9907088]

\bibitem{dob84}
J.~Dobaczewski, H.~Flocard and J.~Treiner,
Nucl. Phys. {\bf A422}, 103 (1984).

\bibitem{bar82}
J.~Bartel, P.~Quentin, M.~Brack, C.~Guet, and H.-B.~H\r{a}kansson, 
Nucl. Phys. {\bf A386}, 79 (1982).

\bibitem{ter03}
E.~Ter\'an, V.~E.~Oberacker and A.~S.~Umar,
Phys. Rev. C {\bf 67}, 064314 (2003).

\bibitem{sto05}
M.~V.~Stoitsov, J.~Dobaczewski, W.~Nazarewicz, P.~Ring, 
Comp. Phys. Comm. {\bf 167}, 43 (2005).

\bibitem{ber91}
G.~F.~Bertsch, H.~Esbensen, Ann. Phys. {\bf 209}, 327 (1991).

\bibitem{ter95}
J.~Terasaki, P.-H.~Heenen, P.~Bonche, J.~Dobaczewski, H.~Flocard, Nucl. Phys. 
{\bf A593}, 1 (1995).

\bibitem{sat98}
W.~Satu{\l}a, J.~Dobaczewski, and W.~Nazarewicz, Phys. Rev. Lett. {\bf 81}, 3599 (1998).

\bibitem{row70}
D.~J.~Rowe, {\it Nuclear Collective Motion}, (Methuen and Co. Ltd., 1970).

\bibitem{bac75}
S.~O.~B\"ackman, A.~D.~Jackson, and J.~Speth, Phys. Lett. {\bf B56}, 209 (1975).

\bibitem{gia81}
N.~Van Giai and H.~Sagawa, Phys. Lett. {\bf B106}, 379 (1981).

\bibitem{gia98}
N.~Van Giai, Ch.~Stoyanov, V.~V.~Voronov, Phys. Rev. C {\bf 57}, 1204 (1998).

\bibitem{kha02}
E.~Khan, N.~Sandulescu, M.~Grasso, N.~Van Giai, 
Phys. Rev. C {\bf 66}, 024309 (2002).

\bibitem{cha98}
E.~Chabanat, P.~Bonche, P.~Haensel, J.~Meyer, and R.~Schaeffer, 
Nucl. Phys. {\bf A635}, 231 (1998).

\bibitem{bel01}
M.~Bellguic, {\it et al.}, Nucl. Phys. {\bf A682}, 136c (2001).

\bibitem{thi00}
P.~G.~Thirolf, {\it et al.}, Phys. Lett. {\bf B485}, 16 (2000).

\bibitem{bec06}
E.~Becheva, {\it et al.}, Phys. Rev. Lett. {\bf 96}, 012501 (2006).

\bibitem{bla77}
J-.P.~Blaizot and D.~Gogny, Nucl. Phys. {\bf A284}, 429 (1977).

\bibitem{miz07}
K.~Mizuyama, M.~Matsuo and Y.~Serizawa, arXiv:0706.1115.

\bibitem{aud95}
G.~Audi and A.~H.~Wapstra, Nucl. Phys. {\bf A461}, 322 (1995).

\bibitem{sau81}
J.~Sauvage-Letessier, P.~Quentin, H.~Flocard, 
Nucl. Phys. {\bf A370}, 231 (1981).

\bibitem{ben00}
M.~Bender, K.~Rutz, P.-G.~Reinhard, J.~A.~Maruhn, Eur. Phys. J. A {\bf 8}, 59 (2000).

\bibitem{dug01}
T.~Duguet, P.~Bonche, P.-H.~Heenen, Nucl. Phys. {\bf A679}, 427 (2001).

\bibitem{yam01}
M.~Yamagami, K.~Matsuyanagi, M.~Matsuo, Nucl. Phys. {\bf A693}, 579 (2001).

\bibitem{per07}
S.~P\'eru, H.~Goutte, J.~F.~Berger, Nucl. Phys. {\bf A788}, 44c (2007).

\bibitem{pen07}
D.~Pe\~na Arteaga, and P.~Ring, Prog. Part. Nucl. Phys. {\bf 59}, 314 (2007).

\bibitem{yam02}
M.~Yamagami, E.~Khan, and N.~Van Giai, 
in {\it Proceedings of the International Symposium on Frontiers of Collective Motions (CM2002)} 
(World Scientific, Singapore, 2002). 

\bibitem{yos08}
K.~Yoshida, in preparation.

\end{thebibliography}
\end{document}